\documentclass[12pt, centerh1]{article}
\usepackage{amssymb,amsmath,colonequals,color,multirow,natbib}

\usepackage{amssymb}
\usepackage{amsfonts, amsmath, amssymb, marvosym,colonequals}
\usepackage{algorithm}
\usepackage{bm,relsize}
\usepackage{geometry}
\usepackage{lscape}
\usepackage{mathtools}
\usepackage{caption}
\usepackage{theorem}
\usepackage{placeins}
\usepackage{hyperref}
\usepackage{enumerate}

\newcommand{\be}{\begin{equation}}
\newcommand{\ee}{\end{equation}} 
\newcommand{\beq}{\begin{equation*}}
\newcommand{\eeq}{\end{equation*}}
\newcommand{\vecx}{\bm{x}}
\newcommand{\vecX}{\mathbf{X}}

\newcommand{\vecz}{\bm{z}}

\newcommand{\Beta}{\mbox{\boldmath$\beta$}}
\newcommand{\vecvarthet}{\mbox{\boldmath$\vartheta$}}

\newcommand{\load}{\mathbf\Lambda}
\newcommand{\noisev}{\mathbf\Psi}

\newcommand{\vecy}{\mathbf{y}} 
\newcommand{\vecU}{\mathbf{U}}
\newcommand{\vecV}{\mathbf{V}}

\newcommand{\vecmu}{\mbox{\boldmath$\mu$}}
\newcommand{\vecalpha}{\mbox{\boldmath$\alpha$}}
\newcommand{\vecSigma}{\mbox{\boldmath$\Sigma$}}

\newcommand{\vectheta}{\mbox{\boldmath$\theta$}}

\newcommand{\vecepsilon}{\mbox{\boldmath$\epsilon$}}

\newcommand{\vecLambda}{\mbox{\boldmath$\Lambda$}}
\newcommand{\vecPsi}{\mbox{\boldmath$\Psi$}}

\newcommand{\matsig}{\mathbf{\Sigma}}

\theorembodyfont{\upshape}

\begin{document}
\title{Mixtures of Skew-t Factor Analyzers}

\author{Paula M.\ Murray, Ryan~P.~Browne and Paul D.\ McNicholas\thanks{Department of Mathematics \& Statistics, University of Guelph, Guelph, Ontario, N1G 2W1, Canada. E-mail: paul.mcnicholas@uoguelph.ca.}
}
\date{Department of Mathematics \& Statistics, University of Guelph.}

\maketitle

\begin{abstract}
In this paper, we introduce a mixture of skew-$t$ factor analyzers as well as a family of mixture models based thereon. The mixture of skew-$t$ distributions model that we use arises as a limiting case of the mixture of generalized hyperbolic distributions. Like their Gaussian and $t$-distribution analogues, our mixture of skew-$t$ factor analyzers are very well-suited to the model-based clustering of high-dimensional data. Imposing constraints on components of the decomposed covariance parameter results in the development of eight flexible models.  The alternating expectation-conditional maximization algorithm is used for model parameter estimation and the Bayesian information criterion is used for model selection. The models are applied to both real and simulated data, giving superior clustering results compared to a well-established family of Gaussian mixture models.   
\end{abstract}



\section{Introduction}
\label{sec:introduction}

Model-based clustering employs finite mixture models to estimate the group memberships of a given set of unlabelled observations.  A finite mixture model has density of the form 
$f(\vecx\mid\vecvarthet)=\sum_{g=1}^{G}\pi_g f_g(\vecx\mid\vectheta_g)$,
where $\pi_g>0$, with $\sum_{g=1}^{G}\pi_g=1$, are the mixing proportions, $\vecvarthet=(\pi_1,\ldots,\pi_G,\vectheta_1,\ldots,\vectheta_G)$ is the vector of parameters, and $f_g(\vecx\mid\vectheta_g)$ is the $g$th component density. Traditionally, model-based clustering has been performed with the component densities taken to be multivariate Gaussian.  However, there has recently been significant interest in non-Gaussian approaches to mixture model-based clustering \citep[e.g.,][]{peel98,mclachlan07,karlis07,lin09,browne12,lee12,franczak12,vrbik12,morris13a,morris13b}. 

\cite{browne13} introduced a mixture of generalized hyperbolic distributions with model density 
$f(\vecx\mid\vecvarthet)=\sum_{g=1}^{G}\pi_g \xi(\vecx\mid\vecvarthet_g)$,
where
\begin{equation}
\begin{split}
\xi(\vecx\mid\vecvarthet_g)=& \bigg[\frac{\chi_g+\delta(\vecx,\vecmu_g|\vecSigma_g)}{\psi_g+\vecalpha_g'\vecSigma_g^{-1}\vecalpha_g}\bigg]^{(\lambda_g-p/2)/2}
\\\times&\frac{[\psi_g/\chi_g]^{\lambda_g/2}K_{\lambda_g-p/2}\bigg(\sqrt{[\psi_g+\vecalpha_g\vecSigma_g^{-1}\vecalpha_g][\chi_g+\delta(\vecx,\vecmu_g|\vecSigma_g)]}\bigg)}{(2\pi)^{p/2}\mid\mathbf\Sigma_g\mid^{1/2}K_{\lambda_g}(\sqrt{\chi_g\psi_g})\mbox{exp}{(\vecmu_g-\vecx)'\vecSigma_g^{-1}\vecalpha_g}},
\label{eq:ghddensity}
\end{split}
\end{equation}
is the density of the generalized hyperbolic distribution, $\vecvarthet_g=(\lambda_g, \chi_g, \psi_g, \vecmu_g, \vecSigma_g, \vecalpha_g)$ is a vector of parameters, and $\delta(\vecx, \vecmu\mid \vecSigma)=(\vecx-\vecmu)'\vecSigma^{-1}(\vecx-\vecmu)$ is the squared Mahalanobis distance between $\vecx$ and $\vecmu$.
As we will discuss herein (Section~\ref{sec:meth}), the skew-$t$ distribution can be obtained as a special case of the generalized hyperbolic distribution \cite[cf.][]{barndorff01}. 

Several alternative forms of the multivariate skew-$t$ distribution have appeared in the literature \citep[e.g.,][]{sahu2003, branco2001, jones2003, ma2004}. However, the form of the distribution arising from the generalized hyperbolic distribution is chosen for this work because of its particularly attractive form, i.e., because its form is particularly conducive to the development of skew-$t$ factors (cf.\ Section~\ref{sec:meth1}). Furthermore, this form of the multivariate skew-$t$ distribution has computational advantages (cf.\ Section~\ref{sec:meth2}) 
and it has not previously been used for model-based clustering.

In this paper, we use this version of the skew-$t$ distribution to introduce a skew-$t$ analogue of the mixture of factor analyzers model, as well as a family of parsimonious models based thereon. The remainder of this paper is laid out as follows. In Section~\ref{sec:back}, we go through some important background material. Then we introduce our methodology, and illustrate our approach on real and simulated data (Section~\ref{sec:anal}). The paper concludes with a discussion and suggestions for future work (Section~\ref{sec:conc}).

\section{Background}\label{sec:back}

\subsection{Mixtures of Factor Analyzers}
The factor analysis model \citep{spearman04} assumes that a $p$-dimensional vector of observed variables can be modelled by a $q$-dimensional vector of latent factors where $q\ll p$, making it useful in the analysis of high-dimensional data.
We can write the factor analysis model as 
$\vecX=\vecmu+\vecLambda \mathbf{U} + \vecepsilon$,
where $\vecLambda$ is a $p\times q$ matrix of factor loadings, $\mathbf{U}\sim N(\mathbf{0}, \mathbf{I}_q)$ is the vector of factors, and $\vecepsilon \sim \mathcal{N}(\mathbf{0}, \vecPsi)$ with $\vecPsi=$diag$(\psi_1, \psi_2, \ldots, \psi_p)$. Under this model, the marginal distribution of $\vecX$ is $\mathcal{N}(\vecmu,\vecLambda\vecLambda'+\vecPsi)$.

\cite{mcnicholas08} introduced a family of eight parsimonious Gaussian mixture models (PGMM) that are suitable for clustering high dimensional data.  These models include both mixtures of probabilistic principal component analyzers \citep{tipping99} as well as mixtures of factor analyzers models \citep{ghahramani97,mclachlan00a} as special cases. The most general such model is a Gaussian mixture model component covariance structure $\vecSigma_g=\vecLambda_g\vecLambda_g'+\vecPsi_g$. 
By allowing the constraints $\vecLambda_g=\vecLambda$, $\vecPsi_g=\vecPsi$, and the isotropic constraint $\vecPsi_g=\psi_g\mathbf{I}_p$, \cite{mcnicholas08} introduced the eight models which comprise the PGMM family.  In addition to introducing a mixture of skew-$t$ distributions model herein, we also implement a family of mixtures of skew-$t$ factor analyzers that is analogous to the PGMM family (Table~\ref{tab:mstfa}). We refer to this family as parsimonious mixtures of skew-$t$ factor analyzers (PMSTFA).
\begin{table}[htbp]
\centering 
\caption{Nomenclature and covariance structure for the members of the PMSTFA family of models.} \label{tab:mstfa}
{\scriptsize\begin{tabular*}{\textwidth}{@{\extracolsep{\fill}}llllr}
\hline
ID &  Loading Matrix & Error Variance & Isotropic & Free Covariance Parameters\\
\hline
CCC& Constrained & Constrained & Constrained & $[pq -q(q-1)/2]+Gp+G+1$\\
CCU & Constrained & Constrained & Unconstrained & $[pq -q(q-1)/2]+2Gp$\\
CUC & Constrained & Unconstrained & Constrained & $[pq -q(q-1)/2]+Gp+2G$\\
CUU & Constrained & Unconstrained & Unconstrained & $[pq -q(q-1)/2] +Gp$\\
UCC & Unconstrained & Constrained & Constrained & $G[pq -q(q-1)/2] +Gp+G+1$\\
UCU & Unconstrained & Constrained & Unconstrained & $G[pq-q(q-1)/2]+2Gp$\\
UUC & Unconstrained & Unconstrained & Constrained & $G[pq-q(q-1)/2]+Gp+2G$\\
UUU & Unconstrained & Unconstrained & Unconstrained & $G[pq-q(q-1)/2]+2Gp+G$\\
\hline
\end{tabular*}}
\end{table}

\subsection{Generalized Inverse Gaussian Distribution}\label{sec:gig}

We write $Y\backsim\text{GIG}(\psi,\chi,\lambda)$ to indicate that the random variable~$Y$ follows a generalized inverse Gaussian (GIG) distribution \citep{good53,barndorff77,blaesild78,halgreen79,jorgensen82} with parameters $(\psi,\chi,\lambda)$. The density of $Y$ is given by
\begin{equation}\label{gig}
p(y\mid\psi,\chi,\lambda) = \frac{\left({\psi}/{\chi}\right)^{\lambda/2}y^{\lambda-1}}{2K_\lambda\left( \sqrt{\psi \chi}\right)}\exp\left\{-\frac{\psi y+\chi/y}{2}\right\},
\vspace{-0.08in}
\end{equation}
for $y>0$, where $\psi,\chi\in\mathbb{R}^+$, $\lambda\in\mathbb{R}$, and $K_{\lambda}$ is the modified Bessel function of the third kind with index~$\lambda$. The generalized inverse Gaussian distribution has some attractive features, including the tractability of the following expected values:
\begin{equation*}\begin{split}
&\mathbb{E}[Y_i|\vecx_i,Z_{ig}=1] =\sqrt{\frac{\chi}{\psi}} \frac{K_{\lambda+1}(\sqrt{\psi\chi})}{K_\lambda(\sqrt{\psi\chi})},\\
&\mathbb{E}[1/Y_i|\vecx_i,Z_{ig}=1] =\sqrt{\frac{\psi}{\chi}} \frac{K_{\lambda+1}(\sqrt{\psi\chi})}{K_\lambda(\sqrt{\psi\chi})}-\frac{2\lambda}{\chi},\\
&\mathbb{E}[\mbox{log}(Y_i)|\vecx_i,Z_{ig}=1] =\mbox{log}\sqrt{\frac{\chi}{\psi}}+\frac{1}{K_\lambda(\sqrt{\psi\chi})}\frac{\delta}{\delta\lambda}K_\lambda(\sqrt{\psi\chi}),
\end{split}\end{equation*}
where $\psi, \chi \in \mathbb{R}$ and $K_{\lambda}$ is the modified Bessel function of the third kind with index $\lambda$.  

\subsection{AECM Algorithm}
The expectation-maximization (EM) algorithm \citep{dempster77} is commonly used for parameter estimation in model-based clustering.  This iterative algorithm facilitates maximum likelihood estimation in the presence of incomplete data, making it well-suited for clustering problems. The E-step of the algorithm consists of computing the expected value of the complete-data log-likelihood based on the current model parameters and the M-step maximizes this expected value with respect to the model parameters. The EM algorithm iterates between these two steps until some convergence criterion is met. 

The expectation-conditional maximization (ECM) algorithm \citep{meng93} is a variant of the EM algorithm that replaces the M-step by multiple conditional-maximization (CM) steps and is useful when the complete-data log likelihood is relatively complicated.  The alternating expectation-conditional maximization (AECM) algorithm \citep{meng97} is a further extension that allows the source of incomplete data to change between the CM steps. An AECM algorithm is used 
for parameter estimation for members of the PGMM family because there are two sources of missing data: the component membership labels $\mathbf{z}_1,\ldots,\mathbf{z}_n$ and the latent factors $\mathbf{u}_1,\ldots,\mathbf{u}_n$. Note that we define $\mathbf{z}_i=(z_{i1},\ldots,z_{iG})$ such that $z_{ig}=1$ if observation~$i$ is in component~$g$ and $z_{ig}=0$ otherwise, for $i=1,\ldots,n$ and $g=1,\ldots,G$. Similarly, we will use the AECM algorithms for parameter estimation for the models used herein (Section~\ref{sec:meth2}). 

\section{Methodology}\label{sec:meth}

\subsection{A Mixture of Skew-$t$ Factor Analyzers}\label{sec:meth1}

The density of a skew-$t$ distribution can obtained as a limiting case of the generalized hyperbolic distribution by setting $\lambda=-\nu/2$ and $\chi=\nu$, and letting $\psi\to0$.  A $p$-dimensional skew-$t$ random variable $\mathbf{X}$ arising in this way has density
\begin{equation}
\begin{split}
\zeta(\vecx\mid\vecmu,\matsig,\vecalpha,\nu)=&
\bigg[\frac{\nu+\delta(\vecx,\vecmu\mid\vecSigma)}{\vecalpha'\vecSigma^{-1}\vecalpha}\bigg]^{{(-\nu-p)}/{4}}\\
&\qquad\qquad\qquad\times\frac{\nu^{\nu/2}K_{(-\nu-p)/2}\bigg(\sqrt{[\vecalpha\vecSigma^{-1}\vecalpha][\nu+\delta(\vecx,\vecmu\mid\vecSigma)]}\bigg)}{(2\pi)^{p/2}\mid\mathbf\Sigma\mid^{1/2}\Gamma({\nu}/{2})2^{\nu/2-1}\exp\{(\vecmu-\vecx)'\vecSigma^{-1}\vecalpha\}},
\label{eq:skewtdensity}
\end{split}
\end{equation}
where $\vecmu$ is the location, $\matsig$ is the scale matrix, $\vecalpha$ is the skewness, and $\nu$ is the value for degrees of freedom.
We write $\vecX\sim\text{GSt}(\vecmu,\matsig,\vecalpha,\nu)$ to denote that the random variable $\vecX$ follows the skew-$t$ distribution such that it has the density given in \eqref{eq:skewtdensity}. 
Now, $\vecX\sim\text{GSt}(\vecmu,\matsig,\vecalpha,\nu)$ can be obtained through the relationship 
\begin{equation}\label{eqn:model1}
\vecX=\vecmu+Y\vecalpha+\sqrt{Y}\vecV,
\end{equation}
where $\vecV \sim \mathcal{N}(\mathbf{0},\vecSigma)$ and $Y \sim \Gamma^{-1}(\nu/2,\nu/2)$, where $\Gamma^{-1}(\cdot)$ denotes the inverse Gamma distribution.
We have $\vecX\mid(Y=y)\sim\mathcal{N}(\vecmu+y\vecalpha,y\vecSigma)$ and so, from Bayes' theorem, $Y\mid(\vecX=\vecx)\sim\mbox{GIG}(\vecalpha'\vecSigma^{-1}\vecalpha, \nu+\delta(\vecx,\vecmu\mid\vecSigma),-(\nu+p)/2)$. 

Our skew-$t$ factor analysis model arises by setting
\begin{equation}\label{eqn:model2}
\vecV=\load\vecU+\vecepsilon,
\end{equation}
where $\load$ is a matrix of factor loadings, $\mathbf{U}\sim N(\mathbf{0}, \mathbf{I}_q)$ is the vector of factors, and $\vecepsilon \sim \mathcal{N}(\mathbf{0}, \vecPsi)$ with $\vecPsi=$diag$(\psi_1, \psi_2, \ldots, \psi_p)$. From \eqref{eqn:model1} and \eqref{eqn:model2}, we can write our skew-$t$ factor analysis model as
\begin{equation}\label{eqn:model3}
\vecX=\vecmu+Y\vecalpha+\sqrt{Y}(\load\vecU+\vecepsilon).
\end{equation}
The marginal distribution of  $\mathbf{X}$ arising from the model in \eqref{eqn:model3} is $\text{GSt}(\vecmu,\load\load'+\noisev,\vecalpha,\nu)$.
Accordingly, our mixture of skew-$t$ factor analyzers model has density
$$f_{\tiny\text{GSt}}(\vecx\mid\vecvarthet)=\sum_{g=1}^{G}\pi_g\zeta(\vecx\mid\vecmu_g,\load_g\load_g'+\noisev_g,\vecalpha_g,\nu_g),$$
where $\vecvarthet$ again denotes all model parameters.

\subsection{Parameter Estimation}\label{sec:meth2}

Parameter estimation is carried out within the AECM algorithm framework. In addition to the GIG expected value formulae given in Section~\ref{sec:gig}, we need the expected value of the component labels, i.e.,
$$\mathbb{E}[Z_{ig}|\vecx_i] = \frac{\pi_g f(\vecx_i|\vectheta_g)}{\sum^{G}_{h=1}\pi_hf(\vecx_i|\vectheta_h)}=:\hat{z}_{ig},$$
as well as the following conditional expectations, which are similar to those used by \cite{mcnicholas08} and \cite{andrews11a}: 
\begin{equation*}\begin{split}
&\mathbb{E}[Z_{ig}(1/Y_{ig})\vecU_{ig}|\vecx_i,1/y_{ig}]=(1/y_{ig})\Beta_g(\vecx_i-\vecmu_g),\\
&\mathbb{E}[Z_{ig}(1/Y_{ig})\vecU_{ig}\vecU_{ig}'|\vecx_i,1/y_{ig}]=\mathbf{I}_q-\Beta_g\vecLambda_g+(1/y_{ig})\Beta_g(\vecx_i-\vecmu_g)(\vecx_i-\vecmu_g)'\Beta_g',
\end{split}\end{equation*}
where $\Beta_g=\vecLambda_g'(\vecLambda_g\vecLambda+\noisev_g)^{-1}$.

In the first stage of our AECM algorithm, the complete-data consist of the data $\vecx_i$, the group membership labels $z_{ig}$, and the latent $y_{ig}$, for $i=1,\ldots,n$, $g=1,\ldots,G$. Hence, the complete-data log-likelihood at this stage is 
$$l_c(\vecvarthet\mid\vecx,\vecy,\vecz)=\sum^{n}_{i=1}\sum^{G}_{g=1}z_{ig}\bigg[\mbox{log} \pi_g+\mbox{log} \phi(\vecx_i\mid\vecmu_g+y_i\vecalpha_g,y_i(\vecLambda_g\vecLambda_g'+\vecPsi_g))+\mbox{log} h(y_i\mid\nu_g/2,\nu_g/2)\bigg].$$  
By maximizing the expected value of the complete-data log-likelihood, we obtain the parameter updates:
$$\hat{\pi}_g= \frac{n_g}{n}, \qquad \hat{\vecmu}_g=\frac{ 1}{m_g}\sum^{n}_{i=1}\vecx_i( \overline{a}_g b_{ig}-1),
\qquad\text{and}\qquad
\hat{\vecalpha}_g=\frac{1}{m_g}\sum_{i=1}^{n}\vecx_i(b_{ig}-\overline{b}_g),$$
where   $a_{ig} =\mathbb{E}[Y_i|\vecx_i,Z_{ig}=1] $, $b_{ig} =\mathbb{E}[1/Y_i|\vecx_i,Z_{ig}=1] $,
$n_g=\sum^{n}_{i=1}z_{ig}$, $\overline{a}_g= ({1}/{n_g})\sum^{n}_{i=1}z_{ig}a_{ig}$, $ \overline{b}_g= ({1}/{n_g})\sum_{i=1}^{n}z_{ig}b_{ig}$, and $m_g = \sum_{i=1}^{n} \overline{a}_gb_{ig}-n_g$. Note that these are analogous to the updates given by \cite{browne13}. 
The update for degrees of freedom does not exist in closed form and the equation 
$$\mbox{log}\bigg(\frac{\hat{\nu}^\text{new}_g}{2}\bigg)+1-\varphi\bigg(\frac{\hat{\nu}^{\text{new}}_g}{2}\bigg)-\sum^{n}_{i=1}\bigg(z_{ig}\log{a_{ig}}+\frac{1}{a_{ig}}\bigg)=0$$ 
must be solved numerically to obtain the updated value of $\nu_{g}^{\tiny\text{new}}$. 

On the second CM-step, the complete-data consist of the data $\vecx_i$, the group membership labels $z_{ig}$, the latent $y_{ig}$, and the latent factors $u_{ig}$, for $i=1,\ldots,n$ and $g=1,\ldots,G$. The expected value of the resulting complete-data log-likelihood is maximized to find updates for the parameters $\vecLambda_1,\ldots,\vecLambda_G$ and $\vecPsi_1,\ldots,\vecPsi_G$. The exact nature of this complete-data log-likelihood and the resulting parameter estimates will depend on which of the eight models in the PMSTFA family is under consideration. These parameter updates are analogous to those given by \cite{mcnicholas08} in the Gaussian case but now the `sample covariance' matrix $\mathbf{S}_g$ takes the more complicated form 
$$\hat{\mathbf{S}}_g=\frac{1}{n_g}\sum_{i=1}^{n}z_{ig}b_{ig}(\vecx_i-\hat{\vecmu}_g)(\vecx_i-\hat{\vecmu}_g)'-\hat{\vecalpha}_g(\bar{\vecx}_g-\hat{\vecmu}_g)'-(\bar{\vecx}_g-\hat{\vecmu}_g)(\hat{\vecalpha}_g)'+A_g\hat{\vecalpha}_g(\hat{\vecalpha}_g)'.$$

Note that the form of the skew-$t$ distribution introduced in Section~\ref{sec:meth1} has a computational simplicity not present in other versions of the skew-$t$ distribution that have been used for model-based clustering \citep[cf.][]{lee13}. The elegant parameter estimation for our skew-$t$ distribution arises because of the properties of the GIG distribution (Section~\ref{sec:gig}). Note that, in the same way described by \citep{mcnicholas10a}, the Woodbury identity \citep{woodbury50} can be used to avoid inverting any non-diagonal $p\times p$ matrices. This latter trick is particularly useful with dealing with higher dimensional data.

\subsection{Model Selection}

The Bayesian information criterion \citep[BIC;][]{schwarz78} is used to select the best model in terms of number of mixture components, number of latent factors, and the covariance structure for each member of the PMSTFA family.  The BIC is defined as BIC$=$2$l(\vecx,\hat{\vecvarthet})- \rho$ log $n$, where $l(\vecx,\hat{\vecvarthet})$ is the maximized log-likelihood, $\hat{\vecvarthet}$ is the maximum likelihood estimate of the model parameters $\vecvarthet$, $\rho$ is the number of free parameters in the model, and $n$ is the number of observations. Support for the use of the BIC in mixture model selection is given by \cite{campbell97} and \cite{dasgupta98}, while \cite{lopes04} provide support for its use in selecting the number of latent factors.

\section{Illustrations}\label{sec:anal}

\subsection{Performance Assessment}

Although all of our illustrations are carried out as \textit{bona fide} cluster analysis, i.e., no knowledge of labels is assumed, the true labels are known and can be used to asses the classification performance of our PMSTFA models. The adjusted Rand index \citep[ARI;][]{hubert85} was used to assess model performance in this study.  The ARI indicates class agreement between the true and estimated group memberships and in doing so, accounts for the fact that random classification would almost certainly classify some observations correctly by chance.  An ARI value of 1 indicates perfect classification, a value of 0 would be expected under random classification, and a negative ARI value indicates classification which is in some sense worse than one would expect under random classification.  

\subsection{Initialization}

For all analyses performed herein, $k$-means clustering was performed on the data and the resulting cluster memberships were used to initialize the $\hat{z}_{ig}$.  The degrees of freedom were initialized at $\nu_g=50$ and the skewness parameters were initialized to be close to zero.  The estimated group means $\hat{\vecmu}_g$ and the sample covariance matrices $\mathbf{S}_g$ were initialized based on the initial $\hat{z}_{ig}$ values, and the matrices $\hat{\vecLambda}_g$ and $\hat{\vecPsi}_g$ were initialized following \cite{mcnicholas08}.  

\subsection{Simulation Study}

\begin{figure}[hbt]
\centering
  \includegraphics[height=2.475in,width=0.49\linewidth]{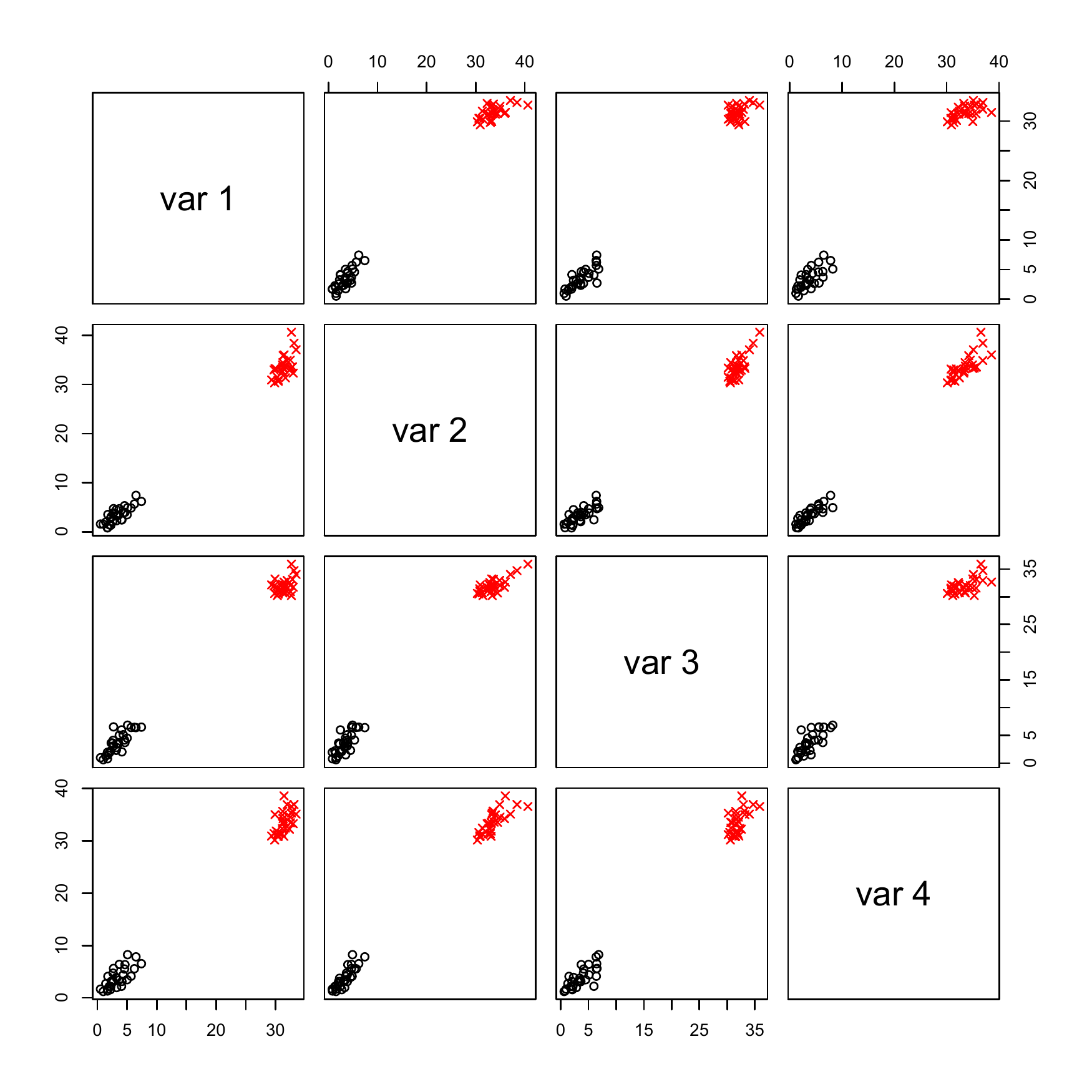} \
  \includegraphics[height=2.475in,width=0.49\linewidth]{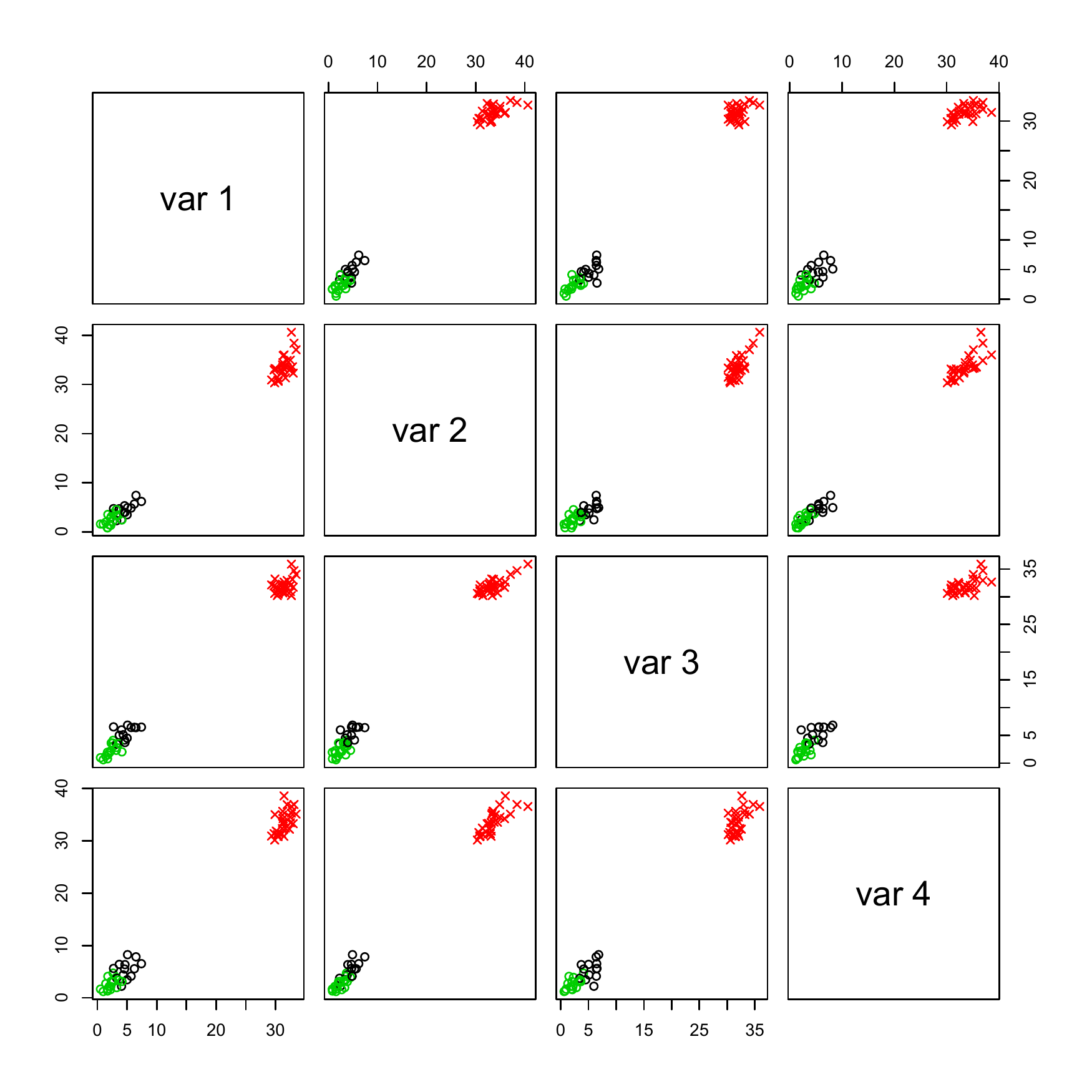}
  \captionof{figure}{Pairs plots illustrating clustering results for the best PMSTFA (left) and PGMM (right) models on four representative variables from the simulated data, where symbols represent true class (i.e., site) and colours reflect predicted classifications.}
  \label{fig:sim}
\end{figure}

A thirteen-dimensional toy data set was simulated from a two-component skew-$t$ mixture model, with $n_1=n_2=30$ and $\vecalpha=(30\quad 30)$. The two components are very well separated but are non-elliptical (cf.\ Figure~\ref{fig:sim}). The PMSTFA models were applied to these simulated data for $G=1,\ldots,4$ groups and $q=1,\ldots,5$ factors using $k$-means starting values.  The model with the highest BIC ($-2431.31$) was the CCC model with $G=2$ groups and $q=1$ latent factors. This model obtained perfect clustering results (AR1=1.00).

\begin{figure}[hbt]
\centering
  \includegraphics[width=0.5\linewidth]{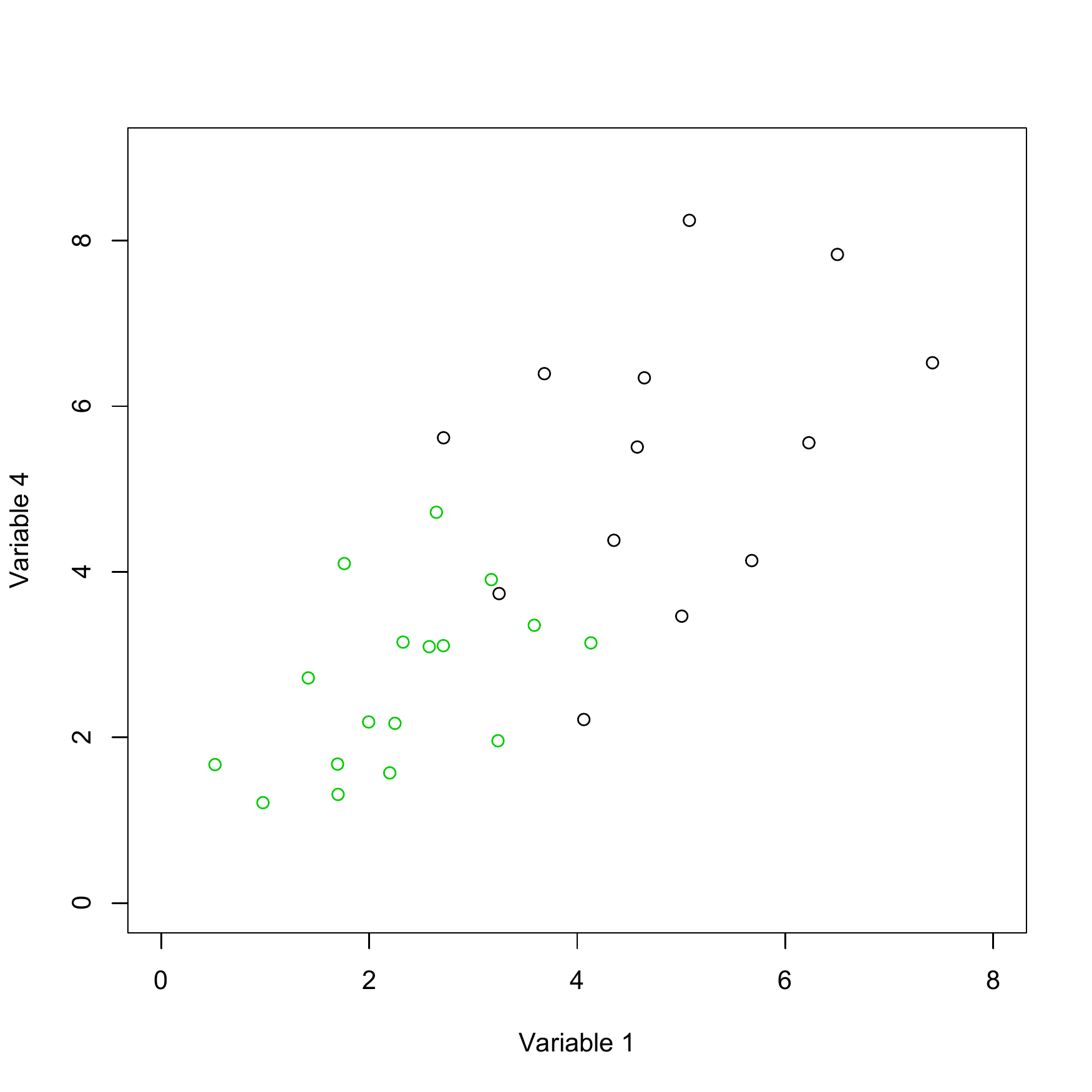}
  \captionof{figure}{Scatter plot illustrating the clustering results for one component of the PGMM best model on one representative variable from the simulated data, where colours reflect predicted classifications and are consistent with Figure~\ref{fig:sim}.}
  \label{fig:sim2}
\end{figure}

For comparison, the PGMM family was also fit to the simulated data using the {\tt pgmm} package \citep{mcnicholas11} for the {\sf R} software \citep{R13}.  The model with the highest BIC ($-2483.46$) was the CUC model with $G=3$ components and $q=1$ latent factor (ARI=0.749). Note that the classification results obtained by this model (Table~\ref{tab:pgmmsim}) could be considered correct if the second and third components were merged (cf.\ Figure~\ref{fig:sim}). This nicely illustrates that Gaussian mixtures --- in this case, a special case of a mixture of factor analyzers model --- can sometimes capture non-elliptical components via \textit{a~posteriori} merging of components (cf.\ Figure~\ref{fig:sim2}).
\begin{table}[h]
\caption{A cross-tabulation of true and predicted group memberships for the selected PGMM model.}
\centering
\begin{tabular*}{0.9\textwidth}{@{\extracolsep{\fill}}lccr}
\hline
 &&Predicted\\
\cline{2-4}	
& 1 & 2 & 3\\ 
  \hline  
True\ 1 & 30 & 0 & 0 \\
True\ 2 & 0 & 17 & 13 \\
\hline
\end{tabular*}
\label{tab:pgmmsim}
\end{table}

\subsection{Australian Institute of Sport Data}

The Australian Institute of Sport (AIS) data contains measured variables on 102 male and 100 female athletes. We consider body fat percentage and body mass index (BMI), taking gender to be unknown.  
The PMSTFA models were fit to the AIS data for $G=1,\ldots,5$ with the number of latent factors fixed at $q=1$.  The model with the highest BIC ($-2224.33$) was the $G=2$ component CCC model (ARI=0.811).  The classification results for this model are given in Table \ref{tab:skewais}.
\begin{table}[h]
\caption{Cross-tabulations of true and predicted group memberships for the selected PMSTFA and PGMM models, respectively, for the AIS data.}
\centering
\begin{tabular*}{0.9\textwidth}{@{\extracolsep{\fill}}lllllll}
\hline
 &\multicolumn{2}{c}{PMSTFA}&&\multicolumn{3}{c}{PGMM}\\
\cline{2-3}	\cline{5-7}
& 1 & 2 && 1 & 2 & 3\\ 
  \hline  
Male & 97 & 5 && 82 & 16&4\\
Female & 5& 95 & & 1 & 9& 90\\
\hline
\end{tabular*}
\label{tab:skewais}
\end{table}

\begin{figure}[hbt]
\centering
  \includegraphics[width=0.48\linewidth]{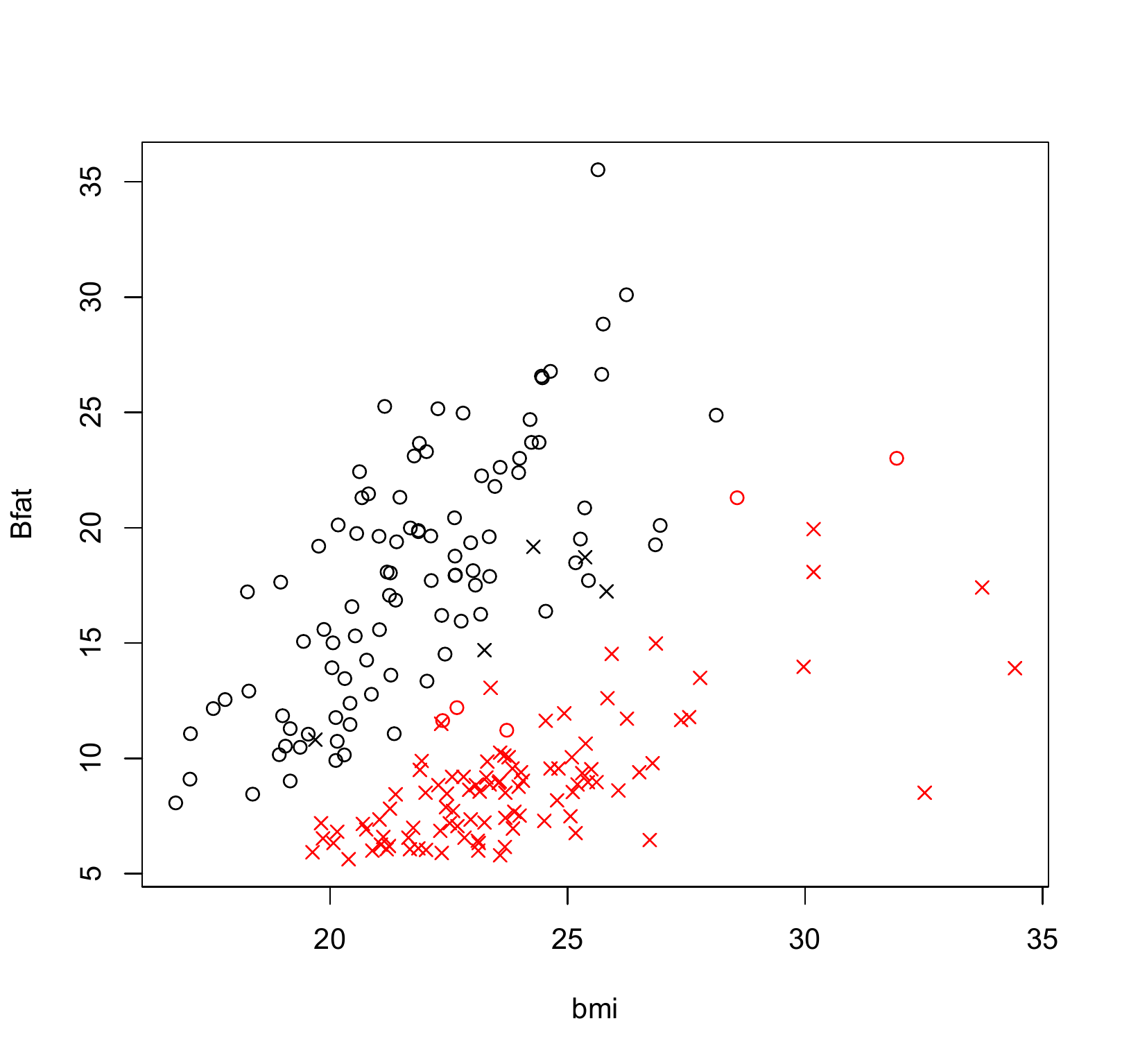} \quad
  \includegraphics[width=0.48\linewidth]{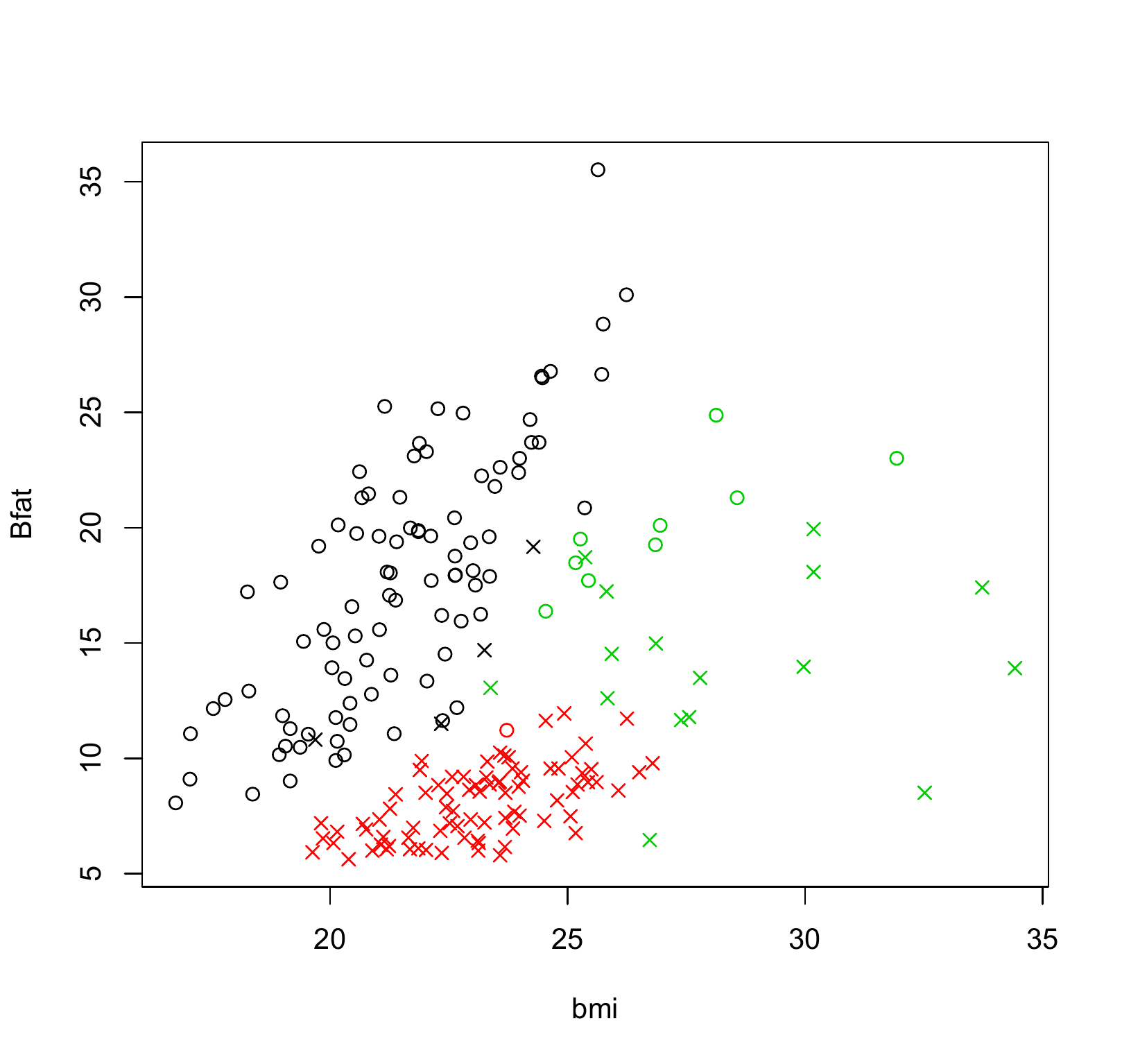}
  \captionof{figure}{Scatter plots illustrating clustering results for the best PMSTFA (left) and PGMM (right) models on the AIS data, where symbols represent true class (i.e., gender) and colours reflect predicted classifications.}
  \label{fig:ais}
\end{figure}

The PGMM models were also fit to the AIS data.  The model with the highest value of the BIC ($-2234.55$) was the $G=3$ component UUC model (ARI=0.685). Comparing the true and predicted classifications (Table~\ref{tab:skewais}), we see that merging components still results in inferior classification performance when compared to the results from the best PMSTFA model (Table~\ref{tab:skewais}). The scatter plots in Figure~\ref{fig:ais} reinforce this point; here, we can see that the best PGMM model has effectively fitted a noise component (i.e., the green component) to pick up points that do not neatly fit into one of the other components. While we have previously demonstrated (Figure~\ref{fig:sim2}) that Gaussian mixtures can sensibly use multiple components to model a single asymmetric cluster, a different situation has emerged here. The PGMM solution on our AIS example illustrates that the imposition of a Gaussian component density is \textit{ipso facto} a stringent constraint that can lead to classification results that are not sensible in any practical sense.

Note that this real data example was chosen to illustrate our PMSFTA models outperforming their Gaussian analogues in a fashion that cannot be overcome by merging Gaussian components. Our chosen model, using $q=1$ latent factor, has same number of misclassifications obtained by \cite{lee12} using mixtures of skew-$t$ distributions. As reported by \cite{vrbik12}, this number may decrease if deterministic annealing \citep{zhou09} is used.

\subsection{Yeast Data}

The PMSTFA models were applied to data containing information on cellular localization sites of 1,484 yeast proteins. The data contains measured variables including the McGeoch's method for signal sequence recognition (mcg), the score of the ALOM membrane spanning region prediction program (alm), and the score of discriminant analysis of amino acid content of  vacuolar and extracellular proteins (vac). The data are available in the UCI Machine Learning Repository with a detailed description of the data given by  \cite{nakai91,nakai92}. For this analysis, we considered two localization sites, CYT and ME3, and clustered the data based on the three variables described above.  

\begin{figure}[hbt]
\centering
  \includegraphics[height=2.475in,width=0.49\linewidth]{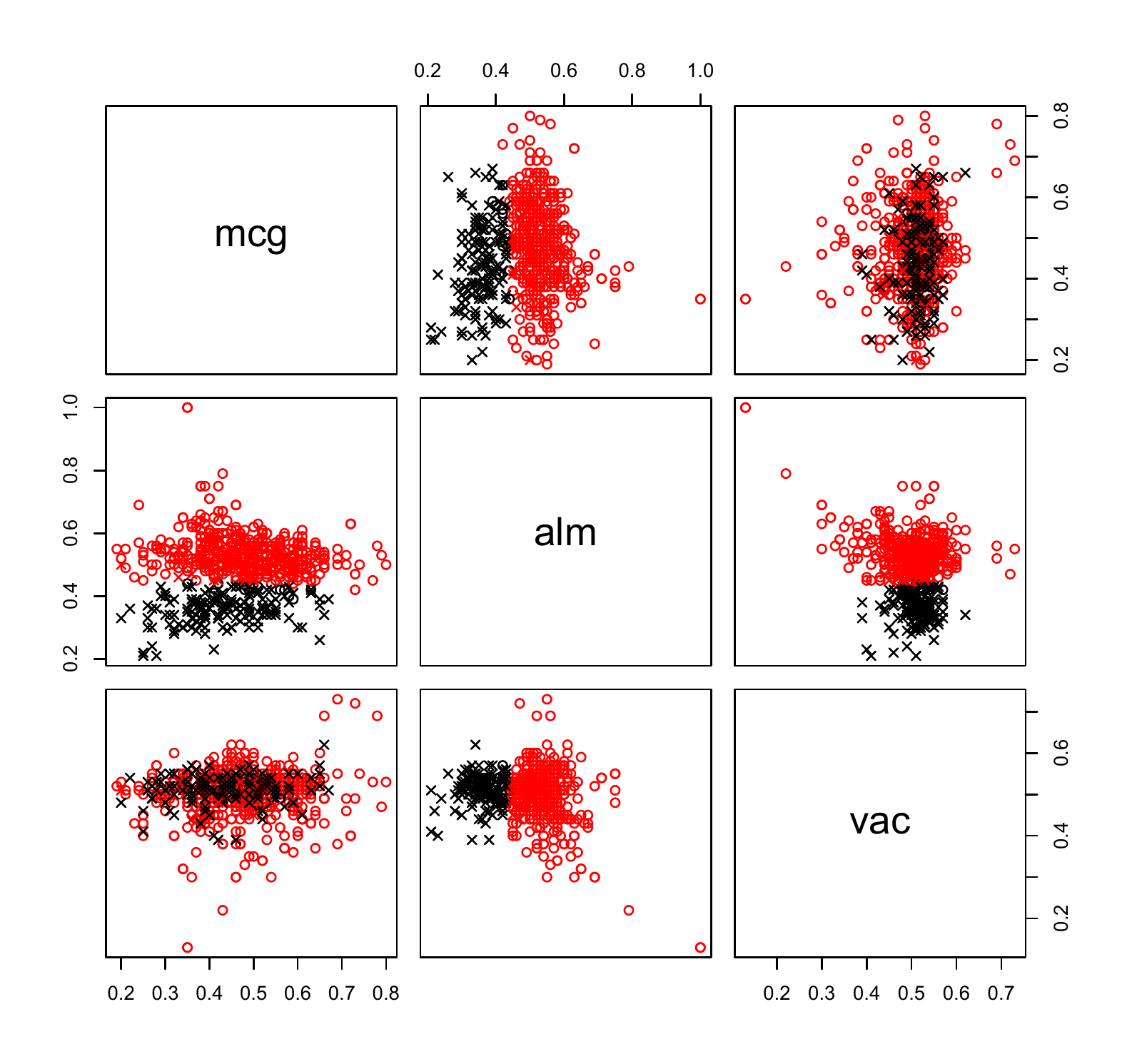} \
  \includegraphics[height=2.475in,width=0.49\linewidth]{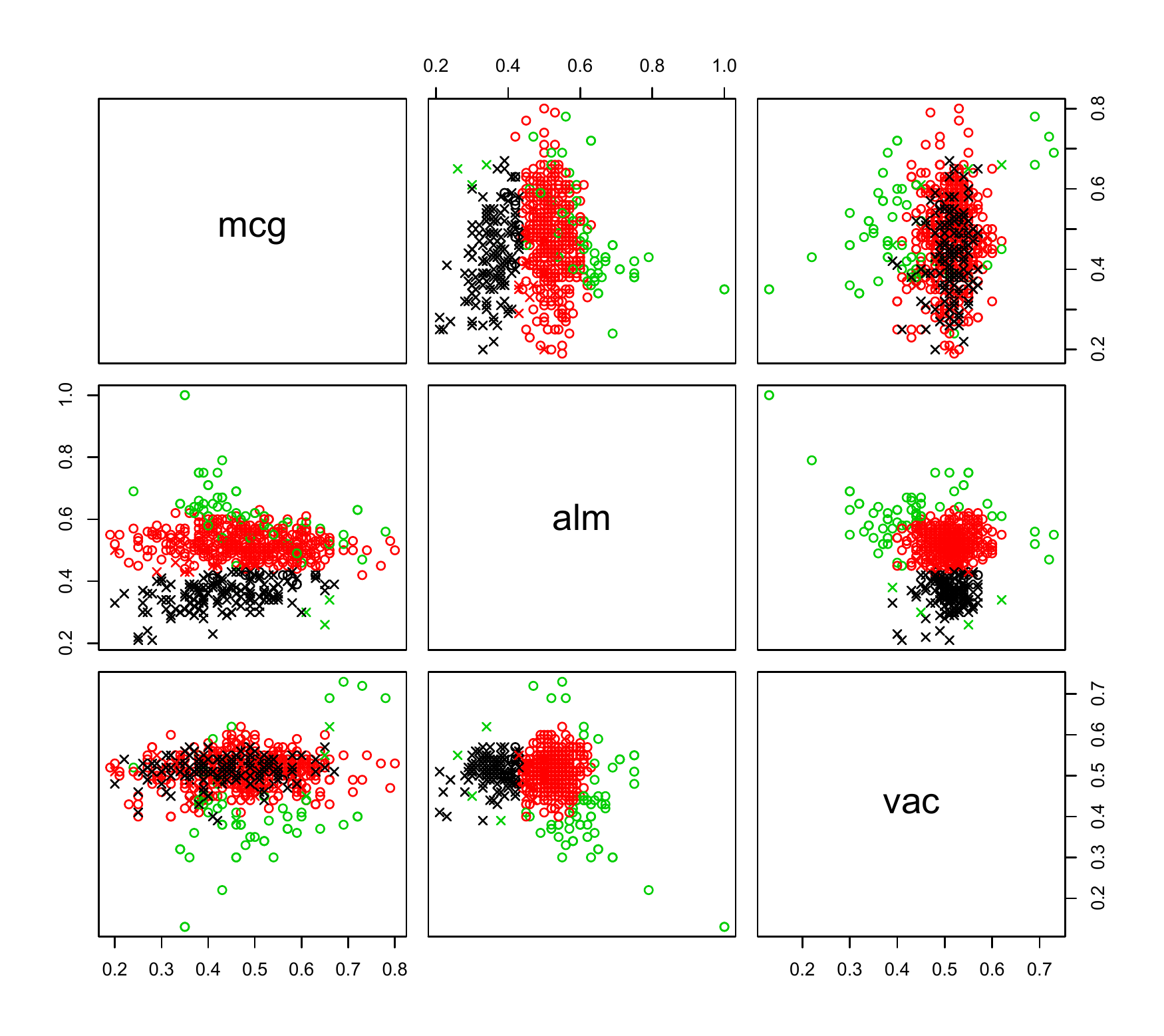}
  \captionof{figure}{Pairs plots illustrating clustering results for the best PMSTFA (left) and PGMM (right) models on the yeast data, where symbols represent true class (i.e., site) and colours reflect predicted classifications.}
  \label{fig:yeast}
\end{figure}

The PMSTFA models were applied for $G=1,\ldots,5$ groups with $q=1$ latent factors from $k$-means starting values.  The best model in terms of BIC ($4226.78$) was the $G=2$ component model ($\text{ARI}=0.85$). The PGMM models were also fit to the data. The model with the highest BIC ($4213.12$) was the $G$=3 component UUC model ($\text{ARI}=0.6$). Cross-tabulations of the true and estimated classifications for both the PMSTFA and PGMM models are given in Table~\ref{tab:yeast}. Note that for the yeast data analysis, as for the AIS data analysis, the best possible merging of Gaussian components still fails to outperform our PMSTFA models (cf.\ Table~\ref{tab:yeast} and Figure~\ref{fig:yeast}).
\begin{table}[h]
\caption{Cross-tabulations of true and predicted group memberships for the selected PMSTFA and PGMM models, respectively, for the yeast data.}
\centering
\begin{tabular*}{0.9\textwidth}{@{\extracolsep{\fill}}lllllll}
\hline
 &\multicolumn{2}{c}{PMSTFA}&&\multicolumn{3}{c}{PGMM}\\
\cline{2-3}	\cline{5-7}
& 1 & 2 && 1 & 2 & 3\\ 
  \hline  
CYT &453& 10 && 391& 64 & 8\\
ME3 &  13& 150 & & 16 & 4 & 143\\
\hline
\end{tabular*}
\label{tab:yeast}
\end{table}

\section{Conclusion}\label{sec:conc}

We have introduced a mixture of skew-$t$ factor analyzers model as well as a family of mixture models based thereon, i.e., the PMSTFA family. This is the first use of a mixture of skewed factor analyzers within the literature. These models were developed using a form of the skew-$t$ distribution that had not previously been used for model-based clustering. This form of the skew-$t$ distribution arises as a special case of the generalized hyperbolic distribution and offers a natural representation for skew-$t$ factor analyzers. Furthermore, borrowing attractive features of the GIG distribution, the form of the skew-$t$ distribution we use lends itself to elegant parameter estimation via an AECM algorithm. 

We illustrated our PMSTFA models on real and simulated data, comparing them to their Gaussian analogous. The real data examples illustrate our PMSTFA family outperforming the PGMM family in ways that cannot be mitigated by merging Gaussian components. Note that the data we used to illustrate our PMSFTA models were low dimensional; however, our models are well suited to high dimensional applications by dint of the fact that the number of covariance parameters is linear in data dimensionality for all eight models. Developing a mixtures of skew-$t$ distributions approach to the analysis of longitudinal data, after the fashion of \cite{mcnicholas10b} and \cite{mcnicholas12}, is a subject of ongoing work. Finally, although used for clustering herein, our PMSTFA models can also be used for model-based classification \citep[see][for the PGMM analogue]{mcnicholas10c} or discriminant analysis \citep{hastie96}. 

\end{document}